\documentstyle[12pt]{article}
\newcommand{\be}{\begin{eqnarray}}
\newcommand{\ee}{\end{eqnarray}}

\begin{document}

\begin{tabbing}
\`SUNY-NTG-94-59\\
\`Dec. 1994
\end{tabbing}
\setlength{\parskip}{20pt}
\vbox to  0.8in{}

\centerline{\Large\bf Hydrodynamics near the QCD Phase Transition:}
\vskip 0.3cm
\centerline{\Large\bf Looking for the Longest-Lived Fireball }
\vskip 2.5cm
\centerline{\large  C.M. Hung and E.V. Shuryak }
\vskip .3cm
\centerline{Department of Physics}\centerline{State University of New York at
Stony Brook}
\centerline{Stony Brook, New York 11794}
\vskip 0.35in
\centerline{\bf Abstract}
We propose a new strategy for  the experimental search of the
QCD phase transition in heavy ion collisions:
One may tune collision energy around the point where
the lifetime of the fireball is expected to be longest.
We demonstrate that the hydrodynamic evolution of excited
nuclear matter does change dramatically as the initial energy density
goes through the ``softest point''
(where the pressure to energy density ratio reaches its minimum). For
our choice of equation of state, this corresponds to $\epsilon_i\approx 1.5
{\rm GeV/fm^3}$ and collision energy $E_{LAB}/A\sim 30$ GeV (for Au+Au).
Various observables seem to show distinct changes near the softest point.
\indent

\eject

  The main objective of experiments related with high energy heavy ion
collisions is the production of sufficiently long-lived and well-equilibrated
multi-particle systems. This allows a macroscopic
description of the system, thus providing information about
the Equation of State (EOS). More specific aims are
to locate the QCD phase
transition and to study a
new state of matter, the quark-gluon plasma.
One obvious way to do this is to reach
very high  energies, such as at RHIC and LHC, so that the  plasma
is produced well above the critical region and may reveal
itself by  outshinning the backgrounds.

  Meanwhile, related experiments are performed at Brookhaven AGS
(10-14 GeV/A)
and CERN SPS (200 GeV/A), and one of the issues
is the magnitude of transverse flow.
 As noticed long ago \cite{SZ,transhydro},
 near the QCD phase
transition the EOS is especially $soft$, leading to
a significant reduction of {\it transverse} expansion (as
 compared to what it would be without phase transition,
e.g. for ideal pion gas EOS). Available
AGS/SPS data have confirmed this
prediction: in particular, no significant growth of the average transverse
momentum $<p_t>$ with the
 multiplicity (or transverse energy)
is observed. Better $\pi\pi$ interferometry data may help to get
quantitative understanding of the transverse expansion issue: current data
indicates that at higher (SPS) energies it seems
to be even $weaker$ than at lower (AGS) ones. Recent results \cite{flow814}
for $non-central$ collisions have revealed ``reaction plane" effects,
similar to what was done before at BEVALAC: but it is still premature
to conclude whether these data
are useful to restrict the EOS.

  In this letter we propose to look at the problem from a different angle.
The ``softness" of the EOS may  affect not only the transverse, but
 the $longitudinal$ expansion as well, leading in principle to
a longer lifetime of the excited system.
However, this can only happen if
the initial conditions are tuned to a narrow range around
the softest point.
As a result, certain observables should show sharp and specific
dependence on the {\it collision energy } around this point.

  To demonstrate the qualitative consequences of the proposed idea,
 we start with the simplest description possible,  namely   the
non-viscous baryon-free hydrodynamics with Fermi-Landau initial conditions.
We have indeed found  that  the central lifetime of the system increases by
 a factor 2.5 in a relatively narrow interval of collision energies.
Clearly more sophisticated models would modify this number, but
we are  confident that  the main features would persist: The total
 lifetime {\it should have a maximum} near the indicated energy region,
 with multiple observable consequences.

  Turning to theoretical expectations on the EOS,
  many important issues concerning
 the QCD phase transition are not yet resolved
(e.g. we do not know whether at physical values of the quark masses
it is first order transition, or a rapid crossover). However,
it is rather  well established that dramatic changes take place
in a $narrow$  temperature interval $  \Delta T \sim 5 \, MeV <<
T_c $, e.g. the energy density changes by about
an order of magnitude, from $\epsilon_{min}$ to $\epsilon_{max}$.
With a somewhat
    broader use of the terminology, one
may refer to this interval as  a
{\it generalized mixed phase}.
 Although it corresponds to narrow (or zero) intervals in T and p,
 the mixed phase  dominates the space-time evolution of the system
created at AGS/SPS energies.

   The particular EOS to be used below corresponds to zero baryon charge
density, and it interpolates between
the ``virial hadron gas" \cite{resonancegas} (with the speed of sound
$dp/d\epsilon=c_s^2 =
0.19$) and the quark-gluon plasma (where
the bag constant is $B=0.32\ {\rm GeV/fm^3}$).
Phase transition (smoothened for numerical purposes) occurs at $T_c=160$ MeV.
We show this $conventional$ EOS   in an unconventional
 way in Fig.1, eliminating T and plotting the (hydrodynamically relevant) ratio
$p(\epsilon)/\epsilon$ versus $\epsilon$.
It emphasizes the existence of a
minimum at $\epsilon=\epsilon_{max}\approx 1.5\ \
{\rm GeV/fm^3}$,
{\it the softest point}.

Since we consider only central collisions,
our calculations are based on the usual equations of ideal relativistic
hydrodynamics with axial symmetry (2+1 D)
The first-order Lax finite
difference scheme is used to solve the equations numerically.
Energy and entropy
conservation is monitored, and we have reproduced results of several
 earlier works  to ensure that technical aspects are under control.
The major uncertainties of hydrodynamic calculations are the initial
conditions: Keeping them as
simple as possible,
we assume zero initial velocity and {\it thermalization} in a uniform disk with
 longitudinal thickness equal to the Lorentz-contracted nuclear diameter.
The total energy of the hydrodynamical subsystem is taken to be $half$
\cite{corrections}
of $\sqrt{S}$.
The reactions chosen are
central Au+Au collisions at varying energies, from 200 GeV/A (SPS) to
10 GeV/A (AGS).

As the collision energy is varied,
radical changes in space-time evolution are evident:
{\bf(a)} At high energy end, we have
{\it  violent longitudinal
expansion}, similar to the well known scale-invariant solution
\cite{Bjorken}. {\bf(b)} Close to  the ``softest point", we found
a {\it slow burning} regime.
 In Fig.2  we  show the space-time evolution for these two cases.
The two pictures are qualitatively different
in almost every aspect. In particular, the contours of longitudinal
velocity (the short-dashed ones) in case (a) are a set of straight lines,
pointing toward the origin: this implies Hubble-like inertial explosion,
without
acceleration. Case (b) is completely different: the expansion velocity
is zero till the approach of the (slowly
propagating)  burning front (boundary between mixed and hadronic
phase), where acceleration takes place.

Energy dependence of some
 global parameters of space-time evolution is shown
in fig.3 (a).
The {\it maximal lifetime
 of the mixed phase} $\tau_M$
(measured at $z,r_t = 0$)
displays a clear peak, reaching the value
$\tau_M\approx   25\ {\rm{fm/c}}$. (The particular collision energy
at which it happens depends on the specific choice of initial conditions:
but the existence of a relatively narrow peak and
its height is a rather stable feature.)
However, longer lifetime does not automatically imply higher
production yields: the price for it  is $smaller$ spatial volume. This is
demonstrated in the same figure by plotting
the total $4-volume$ occupied by the mixed phase $V_M$: its energy dependence
has only a ``shoulder" at the same point.

  Still, such radical changes in the space-time evolution should have
 observable consequences. We have calculated one and two-body
spectra of secondary hadrons, produced at the decoupling
(or freeze-out) surface:
These results will be reported elsewhere, in an expanded version of this
paper.
We have also
evaluated $\gamma$ and $e^+e^-$ production
and compared our  results with preliminary data
\cite{WA80,CERES} from SPS in a separate letter
 \cite{HS1}, where details of this calculation can be found.
We show here only sample results related with the
{\it energy dependence} of observables. In  Fig.3(b) we plot
the height and the width of
 predicted  rapidity distribution of the $e^+e^-$ pairs
(for simplicity, only $M\ =\ 0.5$ GeV is shown, which is representative for
other M values). Both quantities show a noticeable change in
energy dependence
around the ``softest point" of the EOS.
The large height and the small width of the rapidity distribution
at $E_{LAB}\sim 30$ GeV are
 a direct consequence of the long-lived fireball, radiating at rest.
Similar results follow for photon production.

   Note that experimental measurements of $\gamma,e^+e^-$ production at
$E_{LAB}/A\sim 30$ GeV can be made by the $existing$
 detectors WA80 and CERES, provided
 SPS runs at these energies.
Should the predicted specific
behaviour in the dilepton and photon production  be detected,
it will single out the location of the softest point of the EOS.
  Many other observables should also be modified
in this region. A slow-burning
 fireball, with lifetime twice longer than usual, and with much smaller
size, could be identified by
    $\pi\pi,KK$ interferometry.
Possible modifications of strangeness production
 or $J/\psi$ suppression in this region are also interesting possibilities.

  Summarising, we have suggested a new strategy in the search for
the QCD phase transition. The proposal is to
repeat measurements with $existing$ detectors,
scanning  $downward$ from the nominal SPS collision energy.
Photon and dilepton spectra are shown to be sensitive to
the lifetime,  and they can be used to locate the ``softest point"
of the QCD equation of state.

\par

\newpage
\noindent
{\bf Figure captions:}

\noindent
{\bf Fig.1} The equation of state used, shown as pressure to
energy density ratio $p/\epsilon$ versus $\epsilon$.

\noindent
{\bf Fig.2} Comparison of space-time evolution
 for central  Au+Au collisions at {\bf(a)} 129 GeV/A and {\bf(b)}
 30.9 GeV/A (corresponding to the longest fireball lifetime). Solid lines show
the
contours of fixed energy density at the time-longitudinal coordinate (t,z)
plane, at $r_t=0$.
 (The boundaries between quark (Q), mixed (M) and hadronic (H) phases are
shown by the thicker lines, which correspond to energy densities
$1.5$, $0.31$, and $0.13$ $\rm GeV/fm^3$)
The short-dashed lines are contours of fixed
longitudinal velocity, starting from 0.1c for the left-most contour
and increasing by steps
of 0.1c towards the right.

\noindent
{\bf Fig.3} (a)  Lifetime $\tau_M$
of the mixed phase (dashed line, right scale)  and
the corresponding  space-time volume $V_M$ (solid line, left scale)
(b) The height of the dilepton rapidity distribution
$dN_{e+e-}/dy(M=0.5 {\rm GeV},y=0)$
 (dashed line, right scale)  and its full width at half
maximum (solid line, left scale)
versus the collision energy.


\begin{thebibliography}{99}
\bibitem{corrections}
 Another half is attributed to non-thermalized
 nucleons and
their decay
products: we do not attempt to describe  spectra in the fragmentation regions.
We also ignore effect of the excess baryons at central rapidities, which
is only about 10\% correction at the SPS end
of our energy interval,  but larger at the lower end.
\end{thebibliography}
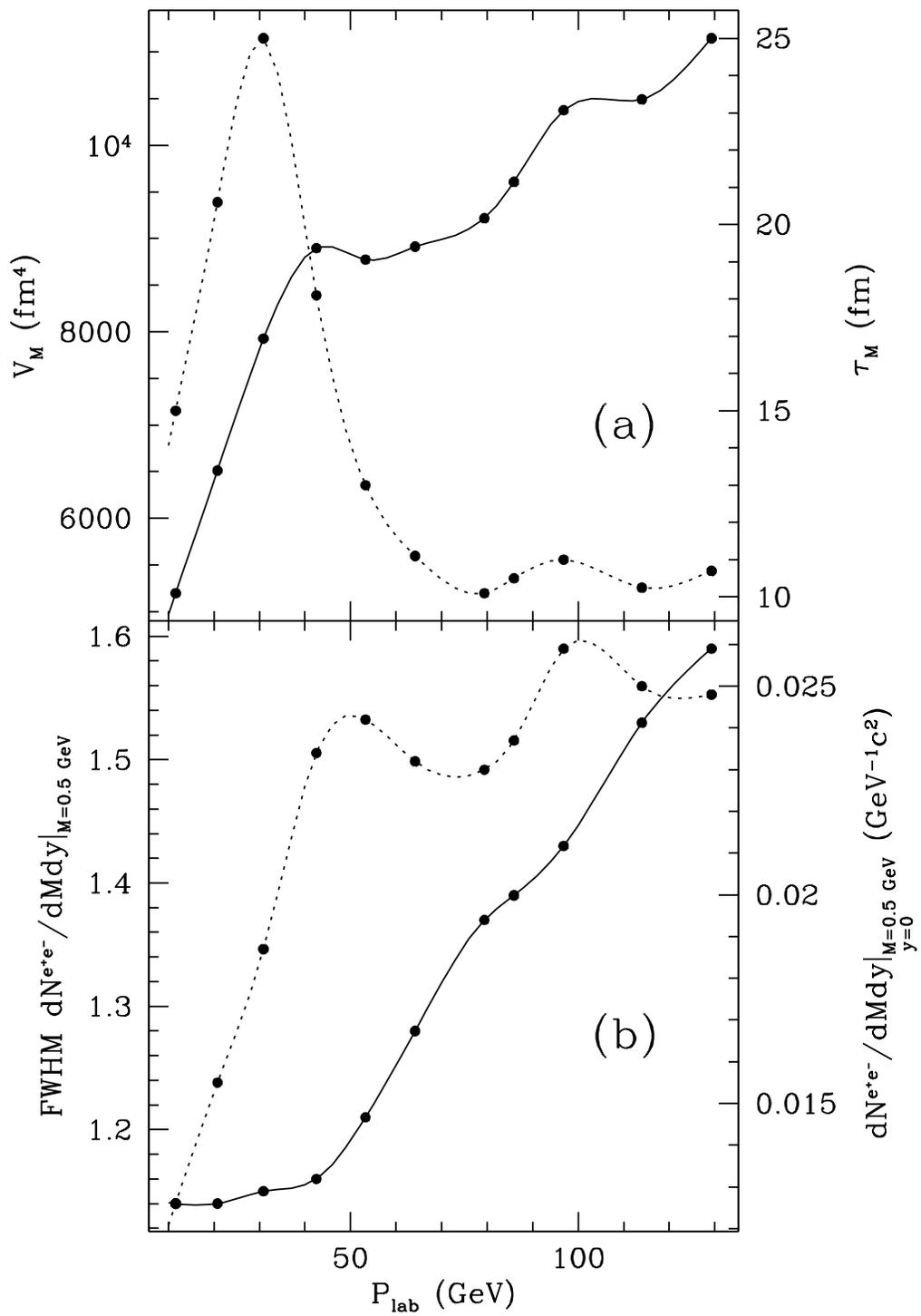
\end{document}